\documentclass[aps,prb,twocolumn,showpacs,10pt,superscriptaddress,floatfix,longbibliography]{revtex4-1}

\usepackage{newtxtext,newtxmath}
\usepackage{graphicx}
\usepackage{subfigure}
\usepackage{amsmath}
\usepackage{amsfonts}
\usepackage{mathrsfs}
\usepackage{bbm}
\usepackage{subfigure}
\usepackage{xcolor}
\usepackage[colorlinks=true,citecolor=blue,linkcolor=blue]{hyperref}
\usepackage{ulem}
\usepackage[T1]{fontenc}

\newcommand{\blue}[1]{\textcolor{blue}{#1}}

\begin{document}

\title{Josephson radiation patterns in underdamped topological Josephson junctions}

\author{Yuejian Zhang}
\affiliation{Department of Physics, Beijing Normal University,
Beijing 100875, China}
\author{Xiping Yang}
\affiliation{Department of Physics, Beijing Normal University,
Beijing 100875, China}
\author{Jia-Jin Feng}
\affiliation{International Center for Quantum Materials, Peking University, Beijing 100871, China}
\author{Zhi Wang}
\email{wangzh356@mail.sysu.edu.cn}
\affiliation{School of Physics, Sun Yat-sen University, Guangzhou 510275, China}
\author{Tianxing Ma}
\email{txma@bnu.edu.cn}
\affiliation{Department of Physics, Beijing Normal University, Beijing 100875, China}
\affiliation{Beijing Computational Science Research Center, Beijing 100193, China}

\begin{abstract}
Josephson radiation is a useful signature for detecting Majorana zero modes in topological superconductors. We study the Josephson radiation of the underdamped topological Josephson junction within the quantum resistively and capacitively shunted junction model. We show that the quantum dynamics of the Majorana two-level system induce oscillatory patterns in the Josephson emission spectra. With the Floquet method, we obtain analytical results for these oscillatory patterns and find that they are well described by Bessel functions. We perform numerical simulations to verify the analytical results and demonstrate that these Bessel radiation patterns exist for a wide range of junction parameters.
\end{abstract}
\date{\today}
\pacs{74.50.+r, 03.65.Sq, 85.25.Dq, 74.78.Na}
\maketitle

%74.50.+r (Tunneling phenomena; Josephson effects),
%03.65.-w (Quantum mechanics),
%03.67.-a (Quantum information),
%74.90.+n (Other topics in superconductivity),
%74.81.Fa (Josephson junction arrays and wire networks),
%76.63.Nm (Quantum wires),
%74.20.Mn (Nonconventional mechanisms),
%85.25.Cp (Josephson devices), 85.25.Dq (SQUIDs)}

\section{Introduction}
In the recent development of quantum computation, a fault-tolerant scheme has been theoretically proposed based on the premises of non-Abelian statistics of quasiparticles\cite{Kitave_2003_quantum_computation}. Such bizarre statistics can appear in superconducting systems for the so-called Majorana zero modes\cite{Kitaev_2001_Majorana,Kwon_2004_Josephson_effect,Sato_2017_opological_RPPhys,Liu_2019_Quadrupole,Zou_2020_symmetries,Liu_2021_monolayer_graphene}, which can construct noise-resistant topological qubits. Braiding Majorana zero modes around one another executes quantum gates whose procession is protected by the topology of the braiding paths\cite{ivanov2001,Alicea_2012_Majorana,Stern_2019_Vortices_PRL,Coulomb_Blockade_Lutchyn_PRL_2019,huang_2021_topological}. Majorana zero modes are theoretically introduced in $p$-wave superconductors and experimentally engineered by putting an $s$-wave superconductor in proximity to a semiconductor with appropriate spin-orbit coupling and Zeeman field
\cite{lutchyn2018review}. Differential conductance is widely used as a signature of Majorana zero modes\cite{lutchyn2018review}. In addition, many recent experiments and theoretical predictions focus on transport and spectroscopic properties of Josephson junctions between two topological superconductors\cite{Fu_2009_JosephsonCurrent_PRB,Ioselevich_2011_PRL,Rokhinson_2012_Josephson_effect,Dominguez_DynamicalDetection_2012_PRB,San_Jose_2013_Andreev_reflection_NewJPhys,Houzet_2013_Dynamics_PRL,Oostinga_2013_HgTe_PRX,Crepin_2014_ParityMeasurement_PRL,Matthews_2014_Dissipation,Sothmann_2016_heat,Peng_2016_Parity,Pico-Cortes_2017_supercurrent_PRB,Cayao_2017_splitting,Dominguez_2017_dynamics_PRB,Sun_2018_QuenchDynamics_PRB,Svetogorov_2020_Critical_PRR,Frombach_2020_poisoning_PRB,Oriekhov_2021_staircase_PRB,Choi_2020_DCShapiro_PRB,Sau_2017_Shapiro_PRB,Majorana_signatures_Lutchyn_PRB_2021}.

In topological Josephson junctions, Majorana zero modes carry a supercurrent which is $4\pi$-periodic in the Josephson phase, known as the fractional Josephson effect
\cite{Kitaev_2001_Majorana}.
Hybridization of Majorana zero modes breaks the exact $4\pi$ periodicity, making the direct transport measurement of $4\pi$ periodicity difficult. Recent experimental efforts are concentrated on studying dynamical properties such as electromagnetic radiation from the junction
\cite{Wiedenmann_2016_Josephson_effect_Nature,Laroche_2019_Josephson_effect_Nature,Deacon_2017_radiation_PRX,Kamata_2018_radiation_PRB,Ren_2019_superconductivity,Murani_2019_Microwave}.
This Josephson radiation was predicted by standard resistively shunted junction model to have a series of spectra with frequencies $f =neV/h$\cite{Kwon_2004_Josephson_effect}, where the first one $f_{\rm M} = eV/h$ corresponds to the $4\pi$-period Josephson effect and the second one $f_{\rm J}=2eV/h$ corresponds to conventional $2\pi$-period Josephson effect. Several recent experiments successfully observed the $4\pi$-period Josephson radiation with frequency $f_{\rm M}$\cite{Deacon_2017_radiation_PRX,Laroche_2019_Josephson_effect_Nature}, but its relation to the Majorana zero modes are still under debate because the signals lack smoking-gun features to be determined uniquely from Majorana zero modes.

\begin{figure}[t]
\begin{center}
\includegraphics[clip = true, width =\columnwidth]{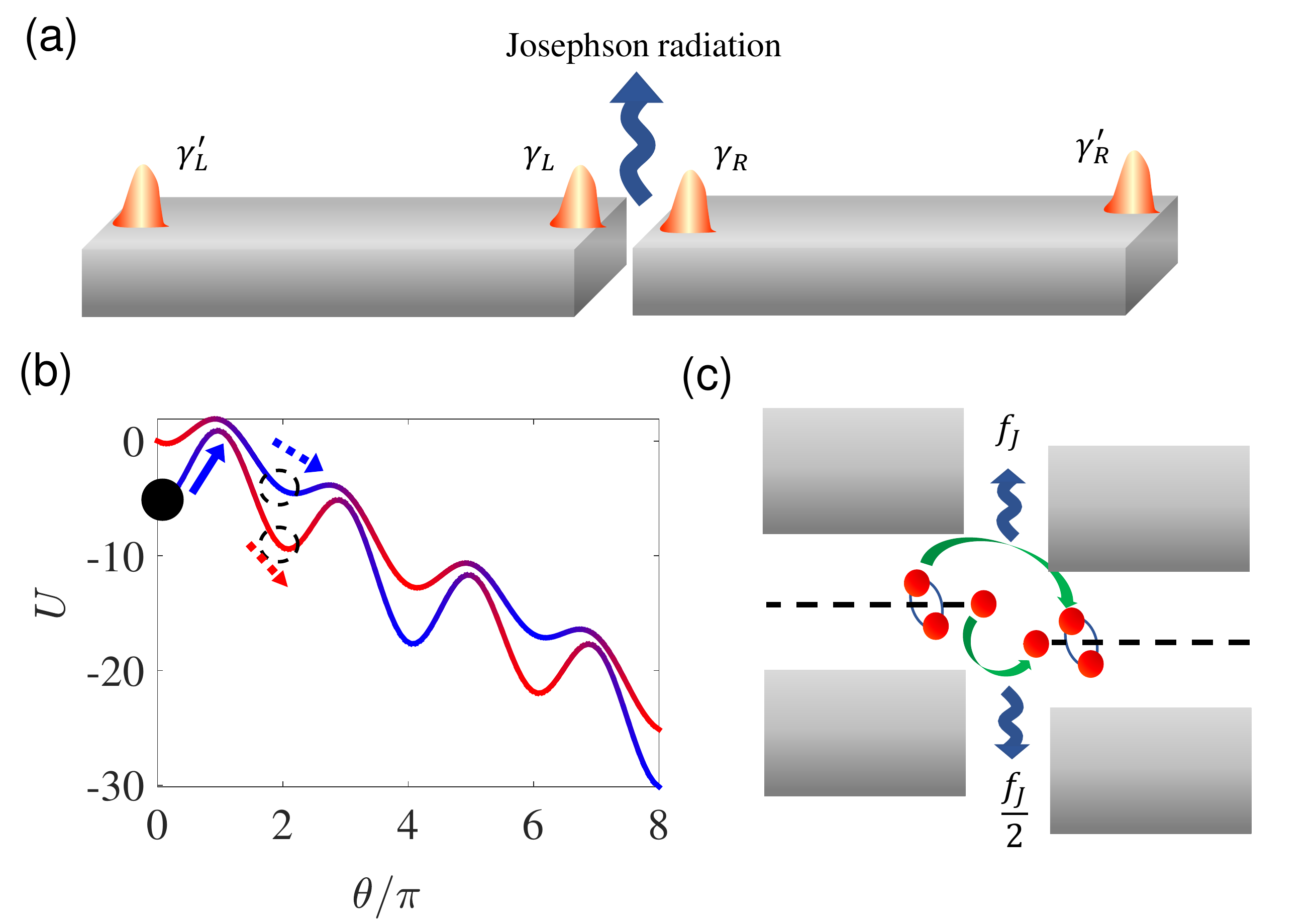}
\caption{~(a) Schematic illustration of a Josephson junction with Majorana zero modes $\gamma_L, \gamma_{L}', \gamma_R, \gamma_{R}'$. Upon current injection above the critical current, the voltage across the junction oscillates and the junction radiates electromagnetic waves. (b) The "tilted-washboard" potential representation of the quantum resistively and capacitively shunted junction model, where the dynamics of the Josephson phase are equivalent to the semi-classical motion of a spin-one-half particle. The potential function for the spin-up state (blue line) and the spin-down state (the red line) have avoided crossing points around $\theta = (2n+1)\pi$. The Landau-Zener transitions would happen at these avoided-crossing points. (c) The semiconductors picture for the Josephson radiation in the voltage-biased-scenario. The coherent Cooper pair tunneling results in a radiation with frequency $f_{\rm J}=2eV/h$, while the coherent single-particle tunneling through the Majorana zero modes contributes radiation with frequency $f_{\rm M}=eV/h$. }
\label{fig:setup}
\end{center}
\end{figure}

Hybridization of Majorana zero modes leads to a quantum two-level system, with its Hamiltonian determined by the Josephson phase\cite{Huang_2015_LZS_PRA}. We have included this Majorana two-level system into the standard theory and built a quantum resistively and capacitively
shunted junction model to study the dynamics of the topological Josephson junction\cite{Feng_2018_dynamics_PRB}. With this model, we have calculated the $I-V$ characteristics curve\cite{Feng_2018_dynamics_PRB} and the Josephson radiation\cite{Feng_2020_Radiation_PRB} of overdamped junctions by neglecting the shunted capacitance, and show results that are consistent with previous experimental reports.
Inspired by recent experimental progress\cite{Wiedenmann_2016_Josephson_effect_Nature,Deacon_2017_radiation_PRX,Kamata_2018_radiation_PRB,Laroche_2019_Josephson_effect_Nature}, we extend the model to study the Josephson radiation of underdamped junctions where the Stewart-McCumber number\cite{tinkham2004} $\beta_c >1$. In this underdamped regime, the shunted capacitance brings more complicate dynamics to the Josephson phase, and the junction potentially exhibits richer emission features in the Josephson radiation.

We consider the Josephson junction with Majorana zero modes as sketched in Fig.~\ref{fig:setup}(a). We numerically simulate the quantum resistively and capacitively
shunted junction model and reveal a unique oscillatory pattern in the spectra, which is due to the quantum oscillation of the Majorana two-level system. For a typical parameter, we analytically solve the equations and find a solution with the form of the Bessel function which satisfactorily captures the main structures of the oscillatory patterns. We further study a wider range of junction parameters and find that the intersection points of the radiation patterns form straight lines that converge to the frequency axis at $f = 4E_{\rm M}/ \pi h$.
These oscillation patterns in the Josephson radiation of the underdamped junction provide additional signatures for confirming the existence of Majorana zero modes.

\section{The quantum resistively and capacitively shunted junction model}\label{II}
The quantum resistively and capacitively shunted junction model was introduced in Ref~[\onlinecite{Feng_2018_dynamics_PRB}]. Within this framework, the dynamics of the Josephson junction are reduced to a set of nonlinear equations, which can be interpreted as equations of motion for a classical particle with spin one-half, subjecting a one-dimensional spin-dependent potential energy as shown in Fig.~\ref{fig:setup}(b). This model contains the least ingredients to describe the dynamics of the Josephson phase and the wave function of the Majorana two-level system. The equations of motion are given by
\begin{eqnarray}\label{eq:RCSJ}
\frac{\hbar C \ddot \theta} {2eI} + \frac{{\rm }\hbar \dot \theta}{2eRI} +I_{\rm 1}\sin{\theta} + I_{\rm 2}\langle i\gamma_{\rm L} \gamma_{\rm R}\rangle \sin{\frac{\theta}{2}}= 1 ,
\end{eqnarray}
where $R$ is the resistance and $C$ is the capacitance of the junction, $\theta$ is the Josephson phase, $I$ is the injected DC current, $I_1 = {I_\mathrm{c1}}/{I}$ with $I_\mathrm{c1}$ the Josephson current from the trivial part of superconductor, $I_2 ={I_\mathrm{c2}}/{I}$ with $I_\mathrm{c2}$ the topological Josephson current contributed by Majorana zero modes, $\langle i\gamma_{\rm L} \gamma_{\rm R}\rangle=|\psi_1|^2-|\psi_0|^2$ and $\psi_{0,1}$ are the two components of the wave function for the two-level system.
The quantum dynamics of the Majorana two-level system is described by the Shcr\"{o}dinger equation with a $\theta$ dependent Hamiltonian,
\begin{eqnarray}\label{eq:linearSE}
i\hbar \frac{{\rm d}}{{\rm d} t} \begin{pmatrix}
\psi_0 \\
\psi_1
\end{pmatrix} = \begin{pmatrix}
E_{\rm M}\cos{\frac{\theta}{2} } & \delta\\
\delta & -E_{\rm M}\cos{\frac{\theta}{2} }
\end{pmatrix} \begin{pmatrix}
\psi_0 \\
\psi_1
\end{pmatrix},
\end{eqnarray}
where $E_{\rm M}$ is the Josephson energy from the coupling of the Majorana zero modes in the junction, and $\delta$ is the hybridization energy from the wave function overlapping of the distant Majorana zero modes on the same side of the junction. To simplify, we replace $\hbar$ with 1 in the following.

The behavior of the model is characterized by the dimensionless Stewart-McCumber number $\beta_c = 2eI_{c1}R^2C/\hbar$. When the capacitance of the junction is negligible so that $\beta_c \ll 1$, the junction is in the so-called overdamped regime. Then the equation is reduced to a first-order difference equation for $\theta$, and the dynamics are largely simplified. This overdamped regime has been explored both experimentally and theoretically\cite{Oostinga_2013_HgTe_PRX,Deacon_2017_radiation_PRX,Feng_2018_dynamics_PRB,Feng_2020_Radiation_PRB}, where nontrivial I-V hysteresis and segmented emission lines were reported and explained. It is interesting to examine the underdamped regime, where the capacitance of the junction is large enough so that $\beta_c > 1$. In this regime, the dynamics of the Josephson phase $\theta$ are much more complicate\cite{tinkham2004}. One would expect to unveil more signatures or patterns for the Majorana zero modes in the transport and spectroscopic signals of the topological junction.

\section{Analytically solutions for the Majorana two-level system}
The quantum resistively and capacitively shunted junction model is described by Eqs. (\ref{eq:RCSJ}) and (\ref{eq:linearSE}) are nonlinear equations that are impossible for analytical exact solutions.
In order to gain insights into the behaviors of the junction in the underdamped regime, we first analyze the dynamics of the Majorana two-level system by taking the lowest-order approximations of the time evolution of the Josephson phase.
When the Josephson junction is injected with a current larger than the critical current, a voltage drop appears and
the superconducting phase difference starts to evolve with time according to the renowned ac Josephson relation\cite{tinkham2004}.
In the present system, this triggers a quantum evolution of the Majorana two-level system.
When the two-level system is dynamically driven across the avoided-crossing points $\theta=(2n+1)\pi$, the quantum state of the Majorana two-level system may either stay in its original parity state or evolve
into the opposite parity state\cite{Huang_2015_LZS_PRA}. This is the Landau-Zener transition\cite{Shevchenko_2010_LZS_Phys.R} as shown in Fig. \ref{fig:setup}(b). The transition probability is given by the
ratio between the energy gap at the avoided-crossing points and the velocity of the Josephson phase.
Experiencing many passages through the avoided-crossing points, the occupation probabilities of the two parity states are governed by the
accumulation of quantum phases acquired at individual passages known as the Landau-Zener-St\"{u}ckelberg interference\cite{Shevchenko_2010_LZS_Phys.R}. The Landau-Zener-St\"{u}ckelberg interference induces an oscillation of the Majorana two-level system
which yields a longer time scale as compared with the oscillation of Josephson phase.

For an applied dc current that is much larger than the critical current, the voltage across the junction has a dominant dc component $V_0$. For the lowest order approximation, we assume that the Josephson phase is purely driven by this dc voltage. Then we have $\theta\approx \omega_{\rm J} t\ (\omega_{\rm J}=2eV_0)$ which can be substituted into the Schr\"{o}dinger equation Eq.~(\ref{eq:linearSE}) to decouple it from the Eq.~(\ref{eq:RCSJ}). After this approximation, the Hamiltonian in Eq.~(\ref{eq:linearSE}) is time-periodic and can be tackled with the Floquet theory, which is the time version of the Bloch theorem. Within the Floquet theory, the original system of two states with time-periodic Hamiltonian in Eq.~(\ref{eq:linearSE}) can be transformed into a system with static Floquet Hamiltonian (see Appendix \ref{appendix:A} for a detailed derivation),
\begin{eqnarray}\label{eq:HF}
{\mathcal{H}}_{\rm f} &&=
\left(\begin{smallmatrix}
\cdot&\cdot&\cdot&\cdot&\cdot&\cdot&\cdot\\%1
\cdot& {\mathcal{H}_0 - \omega_{\rm J} \hat I} & {\mathcal{H}}_1&0& 0&0& \cdot\\
\cdot& {\mathcal{H}}_1 & {\mathcal{H}_0-\frac{ \omega_{\rm J}}{2} \hat I} &{\mathcal{H}}_1& 0&0& \cdot\\
\cdot& 0 & {\mathcal{H}}_1& {\mathcal{H}}_0& {\mathcal{H}}_1&0& \cdot\\
\cdot& 0 & 0&{\mathcal{H}}_1& {\mathcal{H}}_0+\frac{ \omega_{\rm J}}{2} \hat I &{\mathcal{H}}_1& \cdot\\
\cdot& 0 & 0&0& {\mathcal{H}}_1&{\mathcal{H}}_0+ \omega_{\rm J} \hat I & \cdot\\
\cdot&\cdot&\cdot&\cdot&\cdot&\cdot&\cdot\\
\end{smallmatrix}\right),
%\\ \nonumber
\end{eqnarray}
where we have the diagonal and off-diagonal blocks as,
\begin{eqnarray}
{\mathcal{H}}_0 = \begin{pmatrix}
0 & \delta\\
\delta & 0
\end{pmatrix}, \quad
{\mathcal{H}}_{1} =\begin{pmatrix}
E_{\rm M} /2 & 0\\
0 & -E_{\rm M} /2
\end{pmatrix}.
\end{eqnarray}
The diagonal part of the Floquet Hamiltonian is built by infinitely many 2x2 blocks, with each block formed by the time-invariant part of the original Hamiltonian ${\mathcal{H}}_0$ adding energy quanta ${n\omega_{\rm J}}/{2}$
associated with integer number of ``photons"; any two diagonal blocks with one-photon difference are connected by an off-diagonal 2x2 block given by the amplitude of time-periodic part in the
original Hamiltonian.

\begin{figure}[t]
\begin{center}
\includegraphics[clip = true, width =\columnwidth]{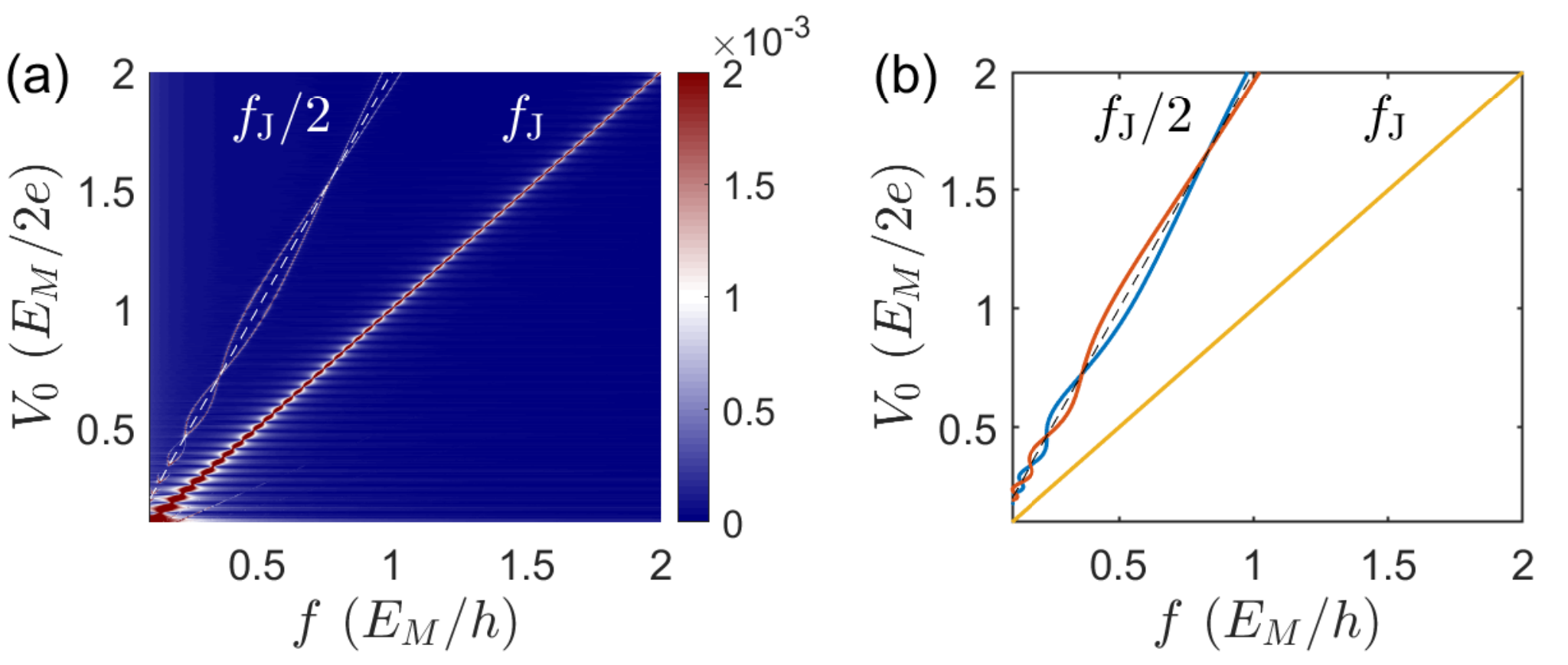}
\caption{ (a) Radiation spectrum for typical underdamped junctions simulated with quantum resistively and capacitively shunted junction model. Junction parameters are taken as $C=5 \times 10^{5}e^2/2E_{\rm M}$, $R=5\times 10^{-3} \hbar/e^2$, $I_{c2}=eE_{\rm M}/\hbar$, $I_{c1} =8eE_{\rm M}/\hbar$, $\delta=0.05E_{\rm M}$. The Stewart-McCumber number is $\beta_c = 100$, making the junction well into the underdamped region. (b) The lowest order oscillation curves of the analytical results shown in Eq.~(\ref{eq:MRadi}). The black dotted line shows the half Josephson frequency $f= f_{\rm J}/2$.}
\label{fig:three}
\end{center}
\end{figure}

In the parameter regions with $\delta \ll \omega_{\rm J}, E_{\rm M}$, the Floquet Hamiltonian ${\mathcal{H}}_{\rm f}$ can be solved with the perturbation method where the Hamiltonian built by ${\mathcal{H}}_0$ blocks is treated as the perturbation term and the rest of ${\mathcal{H}}_{\rm f}$ is treated as the unperturbed Hamiltonian. The unperturbed Hamiltonian can be diagonalized by solving the original time-dependent Schr\"{o}dinger equation Eq.~(\ref{eq:linearSE}) with $\delta$ set to zero, and transforming the solution to the Floquet basis. With this procedure the perturbation term is transformed into a complicated matrix each matrix element takes the form of Bessel functions (see Appendix \ref{appendix:A} for details). The lowest-order perturbation results in the following 2x2 Floquet Hamiltonian,
\begin{eqnarray}\label{eq:HF0}
H_{\rm F}=\begin{pmatrix}
0 & \delta J_{0}(\frac{4E_{\rm M}}{\omega_{\rm J}}) \\
\delta J_{0}(\frac{4E_{\rm M}}{\omega_{\rm J}}) & 0 \\
\end{pmatrix}.
\end{eqnarray}
This effective Floquet Hamiltonian yields a quantum mechanical oscillation for the Majorana two-level system, which is exactly the Landau-Zener-St\"{u}ckelberg interference. As a result, the expectation value of the parity $\langle i\gamma_{\rm L} \gamma_{\rm R}\rangle$ in Eq.~(\ref{eq:RCSJ}) oscillates at a fixed angular frequency (see Appendix \ref{appendix:A} for details),
\begin{eqnarray}\label{eq:LZSfrequencymain}
\omega_{\rm M}=2\delta J_{0}\left(\frac{2E_{\rm M}}{eV_0}\right).
\end{eqnarray}
We notice that the frequency of this Majorana parity state oscillation is linearly proportional to the coupling between the distant Majorana zero modes at each side of the junction. This can be understood intuitively as hybridization between the two parity states.
On the other hand, the dc voltage $V_0$ associated with the increase of superconducting phase difference is involved in a nonlinear way and does not influence much the Majorana
rotation frequency for large voltage. With the increase of the voltage, the reduced probability for parity flipping at a single passage is compensated by the increasing number of passages.

\section{Analytical results for the Josephson Radiation of underdamped junctions}
The quantum oscillation of the Majorana two-level system from the Landau-Zener-St\"{u}ckelberg interference determines the time evolution of $\langle i\gamma_{\rm L} \gamma_{\rm R}\rangle$ term in Eq.~(\ref{eq:RCSJ}) as,
\begin{eqnarray}\label{eq:sew}
\langle i\gamma_{\rm L} \gamma_{\rm R} \rangle \approx s_1 e^{i\omega_{\rm M}t}+s_1^* e^{-i\omega_{\rm M}t},
\end{eqnarray}
where $s_1$ is a constant decided by the initial condition. Plugging this expression back to Eq.~(\ref{eq:RCSJ}), it becomes a self-consistent equation for the Josephson phase, whose solution can provide the information for the Josephson radiation. We consider a dc applied current $I$ and and analyze the time-dependent voltage $V(t)$. Then the Fourier transform of the ac voltage $ V(\omega)$ provides the radiation spectrum. In order to qualitatively understand the behavior of the Josephson radiation, we assume a discrete radiation spectrum of $\omega = \omega_n$, then the time evolution of the voltage is given as $V(t) = V_0+\frac{1}{2}\sum_{n}\left[V_ne^{i\omega_nt}+V_n^*e^{-i\omega_nt}\right]$, where $V_n = V (\omega_n)$ determines the radiation power. With this ansartz, we can obtain the time evolution of the Josephson phase as,
\begin{eqnarray}\label{theta}
\theta=\theta_0+2eV_0t+2e\sum_n\left( \frac{a_n}{\omega_n}\sin\omega_nt-\frac{b_n}{\omega_n}\cos\omega_nt \right),
\end{eqnarray}
where $\theta_0$ is the initial phase, $a_n= (V_n+V_n^*)/2$, $b_n= i(V_n-V_n^*)/2$ are Fourier coefficients of the voltage. Plugging this expression for the Josephson phase into Eq.~(\ref{eq:RCSJ}), we can obtain constraint equations for $V_n$, and a tedious but straightforward calculation (see Appendix \ref{appendix:B} for details) shows that $V_n \neq 0$ only in the case of discrete frequencies at,
\begin{eqnarray}\label{eq:patterns}
\omega_n = p\frac{\omega_{\rm J}}{2}+q \omega_{\rm M} \quad(p,q,\frac{p+q}{2}\in Z, \omega_n>0).
\end{eqnarray}
This result gives the emission lines in the Josephson radiation spectrum. We examine the lowest-order ones by taking $(p,q) = (1,-1), (1,1), (2,0)$, and find the results of $
\omega_1=\frac{\omega_{\rm J}}{2}-\omega_{\rm M}$, $\omega_2=\frac{\omega_{\rm J}}{2}+\omega_{\rm M}$, and $\omega_3=\omega_{\rm J}$. Plugging the Eq. (\ref{eq:LZSfrequencymain}) into these results, we can get the radiation peaks at the frequencies of,
\begin{eqnarray}\label{eq:MRadi}
&&f_1 = \frac{f_{\rm J}}{2}- {\frac{\delta}{\pi} J_{0}\left(\frac{2E_{\rm M}}{eV_0}\right)}, \nonumber \\
&&f_2 = \frac{f_{\rm J}}{2}+ {\frac{\delta}{\pi} J_{0}\left(\frac{2E_{\rm M}}{eV_0}\right)},\nonumber \\
&&f_3= f_{\rm J}.
\end{eqnarray}
We notice that $f_3$ is exactly the radiation at the Josephson frequency which comes from the Cooper pair tunneling process, while $f_1$ and $f_2$ are near to half of the Josephson frequency that comes from the single-electron tunneling process as shown in Fig.~\ref{fig:setup}(c). The surprising part of the Eq.~(\ref{eq:MRadi}) is the two Bessel functions associated with the radiation around the half Josephson frequency. It represents two sets of oscillated emission curves, instead of straight emission lines expected by conventional theories. As shown clearly in Fig.~\ref{fig:three}(b), these oscillatory emission patterns are signatures of the Majorana two-level system oscillation, therefore would be useful for identifying Majorana zero modes in junctions.

The quantum dynamics involving Majorana zero modes, such as the Landau-Zener transitions and the Rabi oscillations have been widely discussed for attempts of achieving quantum manipulations\cite{Coulomb_Blockade_Lutchyn_PRL_2019,Majorana_signatures_Lutchyn_PRB_2021}. The unique feature of the quantum resistively and capacitively shunted junction model is the interplay between these quantum dynamics and the dynamics of the Josephson phase\cite{feng2018hysteresis}. In the following, we reveal that this interplay brings in unique features in Josephson radiation of the underdamped topological junction. 

\begin{figure}[t]
\begin{center}
\includegraphics[clip = true, width =\columnwidth]{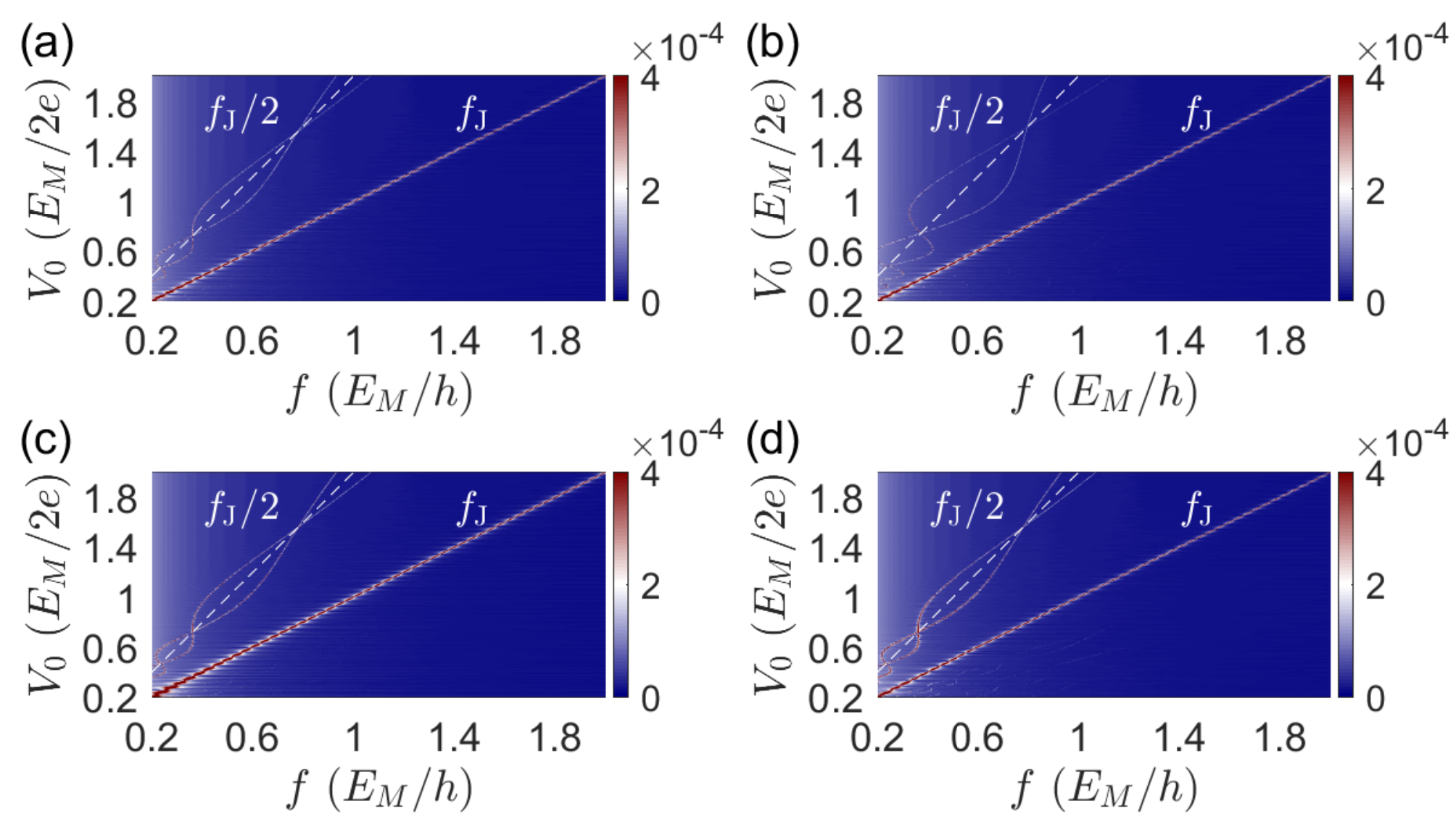}
\caption{Radiation spectrum for underdamped junction simulated with quantum resistively and capacitively shunted junction model with typical parameters. Parameters are taken as (a) $C=2.5 \times 10^{5}e^2/2E_{\rm M}$, $R=5\times 10^{-3} \hbar/e^2$, $I_{c2}=eE_{\rm M}/\hbar$, $I_{c1} =8eE_{\rm M}/\hbar$, $\delta=0.1E_{\rm M}$; (b) $\delta=0.2E_{\rm M}$ with other parameters the same; (c) $C=1.25 \times 10^{5}e^2/2E_{\rm M}$ with other parameters the same; (d) $C=6 \times 10^{4}e^2/2E_{\rm M}$, $R=0.02 \hbar/e^2$, $I_{c1} =2eE_{\rm M}/\hbar$ with other parameters the same.}
\label{fig:S}
\end{center}
\end{figure}

\section{Simulations of underdamped junctions}\label{V}
To verify the results obtained from the analytical method, we perform a direct numerical simulation of the quantum resistively and capacitively shunted junction model. As shown by the Eq.~(\ref{eq:RCSJ}), the Josephson phase begins to increase with an oscillating speed upon injection of a current above the critical value, and the motion of the Josephson phase correlates with the dynamics of the Majorana two-level system\cite{Feng_2018_dynamics_PRB,Feng_2020_Radiation_PRB}. Since the speed of the Josephson phase gives the voltage according to the second Josephson relation $V = \dot \theta/2e$, an oscillating speed of Josephson phase means an oscillating voltage, which, combined with injected dc current, gives electromagnetic radiation, the so-called Josephson radiation. To obtain the radiation spectrum, we first simulate the dynamics of the phase difference and obtain the time evolution of the voltage $V(t)$. From this time-dependent function, we can obtain the dc voltage by time integration $V_0=\int{\rm d}tV(t)$, and obtain the spectrum function through the Fourier transformation,
$V(f)=\int{\rm d}te^{i 2\pi ft} V(t)$.
Since a dc current is injected, this voltage spectrum is exactly the Josephson radiation spectrum. We show the numerical simulation of the Josephson radiation spectrum as a function of the dc voltage and frequency in Fig.~\ref{fig:three}(a), we find one strong emission line at the Josephson frequency $f=f_J$. More importantly, we find two additional emission curves around half of the Josephson frequency $f= f_J/2$. In comparison with the analytical results shown in Fig.~\ref{fig:three}(b), we find that the Josephson radiation of the underdamped topological junction is well described by the Bessel function shape emission patterns.

The results shown in Fig.~\ref{fig:three} are not coincident. The Bessel function emission patterns are
robust in a range of junction parameters, even beyond the parameter regions where the analytical method is applicable. In order to verify this we simulate the coupled equations Eqs.~(\ref{eq:RCSJ}) and (\ref{eq:linearSE}) and demonstrate the radiation spectra for several typical junction parameters in Fig.~\ref{fig:S}.
It is clear that the oscillatory patterns around the half Josephson frequency are robust, while the shape of the patterns may be modified and deviate from the simple Bessel function. These radiation patterns present unique properties of Majorana zero mode in underdamped Josephson junction.

\begin{figure}[t]
\begin{center}
\includegraphics[clip = true, width =\columnwidth]{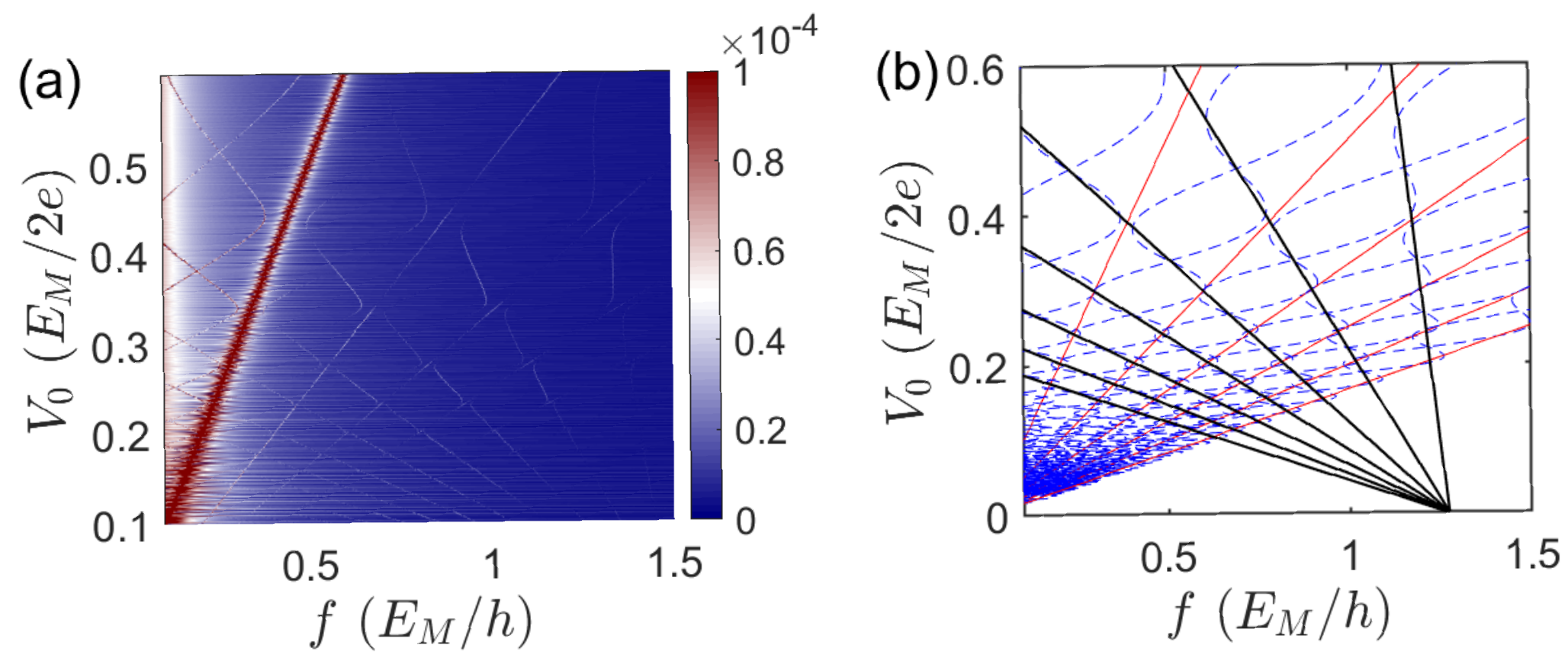}
\caption{(a) Numerical results of the radiation spectrum when the coupling energy Majorana zero modes at one side of the junction is large. Junction parameters are taken the same as in Fig.~\ref{fig:three}(a) except for \blue{$\delta/E_{\rm M}=0.35$}. (b) Analytical results of the oscillation patterns given by Eq.~(\ref{eq:higherbessel}) (blue dotted curve), and the straight lines connecting the intersection points given by Eq.~(\ref{eq:mine}) (black solid line); The Josephson frequencies of $f= n f_{\rm J}/2$ is marked with the red lines. }
\label{fig:four}
\end{center}
\end{figure}

We now examine the analytical results shown in Eq.~(\ref{eq:patterns}) more carefully. The oscillation amplitude of the two Bessel radiation lines linearly depends on $\delta$. In the previous analytical and numerical
calculations, we considered the regime of small $\delta $. In this regime, the oscillation amplitude is smaller compared with the distance between different emission curves with different $(p,q)$. However, in the case of large $\delta$, the oscillation amplitude is large and the oscillatory patterns almost overlap with each other. We show the analytic results of radiation patterns with $(n,\pm 1)$ in Fig.~\ref{fig:four}(b), where the red lines mark the higher-order Josephson frequency at $f=nf_{\rm J}$, and the blue dotted lines are radiation patterns caused by Majorana zero modes approximately described by the Bessel functions
\begin{eqnarray}\label{eq:higherbessel}
f_{\rm M\pm}^n=\frac{(2n-1)eV_0}{2\pi}\pm\frac{\delta}{\pi} J_0( \frac{2E_{\rm M}}{eV_0}).
\end{eqnarray}
We demonstrate our numerical results in this parameter regime in Fig.~\ref{fig:four}(a). Novel radiation characteristics appear in the radiation pattern.
We have an interesting observation in Fig.~\ref{fig:four}(a). The oscillation lines of different orders are connected to form a radiation spectrum with negative slopes. We can analytically explain this feature by revisiting the Eq. (\ref{eq:higherbessel}). The intersections of two $f_{\rm M\pm}^n$ rays of the same order $n$ are zero points of $J_0( 2E_{\rm M}/{eV_0})$. If these points are connected in the way shown by the black solid line in Fig.~\ref{fig:four}(b), we can find that the connected black line overlaps with the radiation line caused by Majorana zero modes. These results predict new radiation lines when $\delta$ is large. To have a quantitative understanding of the result, we approximate the Bessel functions with the trigonometric functions, and we would have the black solid lines in Fig.~\ref{fig:four}(b) expressed as,
\begin{eqnarray}\label{eq:mine}
\frac{f}{(E_{\rm M}/2{\rm \pi})}=-(l+\frac1 4) \frac{V_0}{(E_{\rm M}/2e)}+\frac{4}{\rm \pi},
\end{eqnarray}
where $l$ can be any integer. This analytical result shows that the new radiation lines converge to the frequency axis at $4/{\rm\pi}$ (see Appendix \ref{appendix:B} for a detailed derivation), as shown in Fig.~\ref{fig:four}(b) .

Finally, we discuss the junction parameters taken in the numerical calculations. The Majorana coupling energy $E_{\rm M}$ is taken as the unit for the numerical calculation. For the nanowire junction, it was expected to be in the order of $E_{\rm M}=10{\rm \mu eV}$. With this estimation, we find that the other parameters taken in Fig.~\ref{fig:three}(a) are $I_{c1} \approx 20{\rm nA}$, $I_{c2}\approx 2.5{\rm nA}$, $\delta \approx 0.5{\rm \mu eV}$, $ R \approx 20{\rm \Omega}$, and $C \approx 4 {\rm nF}$ with the charging energy $e^2/2C \approx 2 \times 10^{-5}\mu eV$. These parameters reasonably gives a Stewart-McCumber number of $\beta_c \approx 100$, making the junction well into the underdamped region. 

\section{conclusion}
In summary, we use the quantum resistively and capacitively shunted junction model to study the Josephson radiation of an underdamped topological Josephson junction with Majorana zero modes. We obtain the radiation spectra with both numerical simulation and analytical solution. Unique to our model, we show that the $4\pi$-periodic Josephson emission presents oscillating patterns which are well described by Bessel functions. We reveal that they are coming from the constructive Landau-Zener-St\"{u}ckelberg interference of the Majorana two-level system. In addition, we find that the intersection points of the oscillating patterns form straight lines which converge to the frequency axis. These patterns in the radiation spectra provide features for detecting the Majorana zero modes.
\begin{acknowledgments}
This work was supported by NSFC (Nos. 11974049, 11774033, 11774435, and 12174453), Beijing Natural Science Foundation (No. 1192011), and Guangdong Basic and Applied Basic Research Foundation (No. 2019A1515011620).
We acknowledge the support of HSCC of Beijing Normal University.
\end{acknowledgments}

\appendix

\section{Floquet theory and analytical calculation of $\omega_{\rm M}$}\label{appendix:A}
With the approximation of $\theta\approx\omega_{\rm J}t$, the Hamiltonian in Eq.~(\ref{eq:linearSE}) is time-periodic. In general, the solution can be given in the form
\begin{eqnarray}
\Psi(t) \equiv (\psi_0, \psi_1)^{\rm T}= \tilde \psi(t) e^{-iQt},
\end{eqnarray}
with $\tilde \psi(t)$ a time-periodic function with the same period as the Hamiltonian $\tilde \psi(t)= \tilde \psi(t+{4\pi}/\omega_{\rm J})$, and $Q$ the eigenvalue to be determined.
It is natural to decompose the Hamiltonian, and accordingly the wave function, into Fourier components
\begin{eqnarray}
{\mathcal{H}} = \sum_n {\mathcal{H}}_n e^{i n \omega_{\rm J} t/2}, \hspace{5mm} \tilde \psi(t) = \sum_n \tilde \psi_n e^{i n \omega_{\rm J} t/2},
\end{eqnarray}
with the Hamiltonian components
\begin{eqnarray}
{\mathcal{H}}_0 = \begin{pmatrix}
0 & \delta\\
\delta & 0
\end{pmatrix}, \quad
{\mathcal{H}}_{\pm 1} =\begin{pmatrix}
E_{\rm M} /2 & 0\\
0 & -E_{\rm M} /2
\end{pmatrix},
\end{eqnarray}
and all other matrices are zero in the present case. Plugging them back into Eq.~(\ref{eq:linearSE}), we arrive at the following static secular equation
\begin{eqnarray}
\sum_{l} \left({\mathcal{H}}_{n-l} + \frac{n \omega_{\rm J}}{2} \delta_{nl} \hat I\right) \tilde\psi_l = Q \tilde\psi_n,
\end{eqnarray}
with $\hat I$ the 2x2 identity matrix.
Defining a vector $ \tilde \Psi = (..., \tilde \psi_{-1}, \tilde \psi_0, \tilde \psi_1,...)^{\rm T}$, we obtain the time-independent Floquet Hamiltonian

\begin{eqnarray}
{\mathcal{H}}_{\rm f} &&=
\left(\begin{smallmatrix}
\cdot&\cdot&\cdot&\cdot&\cdot&\cdot&\cdot\\%1
\cdot& {\mathcal{H}_0 - \omega_{\rm J} \hat I} & {\mathcal{H}}_1&0& 0&0& \cdot\\
\cdot& {\mathcal{H}}_1 & {\mathcal{H}_0-\frac{ \omega_{\rm J}}{2} \hat I} &{\mathcal{H}}_1& 0&0& \cdot\\
\cdot& 0 & {\mathcal{H}}_1& {\mathcal{H}}_0& {\mathcal{H}}_1&0& \cdot\\
\cdot& 0 & 0&{\mathcal{H}}_1& {\mathcal{H}}_0+\frac{ \omega_{\rm J}}{2} \hat I &{\mathcal{H}}_1& \cdot\\
\cdot& 0 & 0&0& {\mathcal{H}}_1&{\mathcal{H}}_0+ \omega_{\rm J} \hat I & \cdot\\
\cdot&\cdot&\cdot&\cdot&\cdot&\cdot&\cdot\\
\end{smallmatrix}\right),
%\\ \nonumber
\end{eqnarray}
which is shown as Eq.~(\ref{eq:HF}).

The basis states for the Floquet Hamiltonian ${\mathcal{H}}_{\rm f}$ are the Floquet states $|n,\alpha \rangle$, with $\alpha$ referring to the two states $|S_j\rangle$ ($j=$~0 and 1) and $n$ to the Fourier component (or equivalently the number of photons).
This Floquet Hamiltonian is static and thus can be used to describe the quantum dynamics of the original 2x2 time-periodic Hamiltonian.
Especially the transition probability between two quantum states $|\psi_0 \rangle$ and $|\psi_1 \rangle$
can be expressed as a summation of those between corresponding Floquet states $|0,\psi_0 \rangle$ and $|n, \psi_1 \rangle$
\begin{eqnarray}
T_{|\psi_0 \rangle \rightarrow |\psi_1 \rangle} = \sum_{n} \left|\langle n,\psi_1 |e^{-i{\mathcal{H}}_f (t-t_0)}|0,\psi_0 \rangle \right|^2.
\label{transition}
\end{eqnarray}
Therefore, the problem of evaluating the transition probability between two quantum states is governed by
a time-dependent Hamiltonian is
reduced to a task to solve a time-independent Floquet Hamiltonian, which can be treated by conventional techniques in quantum mechanics.

In the case of $\delta \ll \omega_{\rm J}, E_{\rm M}$, we can divide the Floquet Hamiltonian into two parts
\begin{eqnarray}
{\mathcal{H}}_{\rm f} = H_0 + H',
\end{eqnarray}
with the perturbation term
\begin{eqnarray}\label{eq:H0V}
H'=\delta \sum_{n}\big(|n,\psi_0 \rangle \langle n,\psi_1 |+|n,\psi_1 \rangle \langle n,\psi_0| \big),
\end{eqnarray}
and $H_0$ is the rest of ${\mathcal{H}}_{\rm f}$. The direct diagonalization of the Floquet Hamiltonian $H_0$ is complicate. However, it can be achieved by reconsidering the original time-dependent problem. We notice that, for $\delta=0$, the solution of Eq.~(\ref{eq:linearSE}) can be given directly
\begin{eqnarray}
\psi_0 (t) &=& e^{- 2i \frac{ E_{\rm M}}{\omega_{\rm J}} \sin \frac{\omega_{\rm J} t}{2} } \psi_0 (0), \nonumber\\
\psi_1 (t) &=& e^{ 2i \frac{ E_{\rm M}}{\omega_{\rm J}} \sin \frac{\omega_{\rm J} t}{2} } \psi_1(0).
\end{eqnarray}
Using the Jacobi-Anger expansion\cite{Cuyt_2008_Handbook}
\begin{eqnarray}
e^{ i z \sin \gamma} = \sum^{+\infty}_{n = -\infty}J_n(z)e^{in\gamma},
\end{eqnarray}
where for the present case $z=-{2 E_{\rm M}}/(\omega_{\rm J})$ and $\gamma={\omega_{\rm J} t}/{2}$, the above solution can be decomposed into Fourier components
\begin{eqnarray}
\begin{pmatrix}
\psi_0(t)\\
\psi_1(t)
\end{pmatrix} = \sum^\infty_{n = -\infty}
\begin{pmatrix}
J_n(z) \psi_0(0)\\
J_n(-z) \psi_1(0)
\end{pmatrix}
e^{in\gamma},
\end{eqnarray}
with the form useful for Floquet analysis. With $\psi_0(0) = 0, \psi_1(0) = 1$ and $\psi_0(0) = 1, \psi_1(0) = 0$, we obtain two sets of Floquet eigenfunctions
\begin{eqnarray}
&&\tilde \Psi_{+,0} = (..., J_{-1}(z),0,J_{0}(z),0,J_{1}(z),0,...)^{\rm T}, \\
&&\tilde \Psi_{-,0} = (..., 0,J_{-1}(-z),0,J_{0}(-z),0,J_{1}(-z),...)^{\rm T}, \nonumber
\end{eqnarray}
associated with $Q_0=0$ where the subscript refers to the non-perturbed case.
Due to the translational symmetry of the Floquet Hamiltonian, eigenfunctions associated with eigenvalues $Q_0= n\omega_{\rm J}/2$ are given straightforwardly as
\begin{eqnarray}
&&\tilde \Psi_{+,n} = (..., J_{-n-1}(z),0,J_{-n}(z),0,J_{-n+1}(z),0,...)^{\rm T}, \nonumber \\
&&\tilde \Psi_{-,n} = (..., 0,J_{-n-1}(-z),0,J_{-n}(-z),0,J_{-n+1}(-z),...)^{\rm T}.\nonumber \\
\end{eqnarray}
Using these eigenfunctions as the new basis, one has the kinetic equivalent Hamiltonian $K_{0,{m,n}}=\hat I \delta_{mn}n\omega_{\rm J}/2 $ and
\begin{eqnarray}
K'_{n,m}
&&=\begin{pmatrix}
\langle \tilde \Psi_{+,n} | H' | \tilde \Psi_{+,m} \rangle & \langle \tilde \Psi_{+,n} | H' | \tilde \Psi_{-,m} \rangle \\
\langle \tilde \Psi_{-,n} | H' | \tilde \Psi_{+,m} \rangle & \langle \tilde \Psi_{-,n} | H' | \tilde \Psi_{-,m} \rangle \\
\end{pmatrix}
\\ \nonumber
&&=\delta\begin{pmatrix}
0 & \sum_{i}{J_{i-n}(z)J_{i-m}(-z)} \\
\sum_{i}{J_{i-n}(-z)J_{i-m}(z)} & 0 \\
\end{pmatrix}.
\end{eqnarray}
Exploiting the relations for Bessel functions\cite{Andrews_1999_SpecialFunction}
\begin{equation}
\sum_{i}{J_{i-n}(z)J_{i-m}(-z)} = J_{m-n}(2z),
\end{equation}
we arrive at the kinetic perturbation Hamiltonian
\begin{eqnarray}
K'_{n,m}=\delta\begin{pmatrix}
0 & J_{m-n}(2z) \\
J_{n-m}(2z) & 0 \\
\end{pmatrix}.
\end{eqnarray}
Now the unperturbed part of Hamiltonian $H_0$ is diagonalized, while, at the cost, the interaction part becomes complicated with more non-zero elements as compared to that in Eq.~(\ref{eq:H0V}).
However, for the case $\delta \ll \omega_{\rm J}, E_{\rm M}$, only the transitions between $|0,\psi_0 \rangle$ and $|0,\psi_1 \rangle$ is important in Eq.~(\ref{transition})
to the lowest-order perturbation, which results in the following 2x2 Floquet Hamiltonian as shown in Eq.~(\ref{eq:HF0})
\begin{eqnarray}
H_{\rm F}=\begin{pmatrix}
0 & \delta J_{0}(2z) \\
\delta J_{0}(2z) & 0 \\
\end{pmatrix}.
\end{eqnarray}
The evolution operator for this effective Floquet Hamiltonian is
\begin{eqnarray}
\hat U(t) = e^{-i\hat H_F t}
= e^{-i\delta J_{0}(2z) \tau_x t},
\end{eqnarray}
which yields a quantum mechanical oscillation
\begin{eqnarray}
|\psi_1(t)|^2 =\cos^2 (\omega_{\rm LZS} t),\hspace{3mm} |\psi_0(t)|^2 =\sin^2 (\omega_{\rm LZS} t),\nonumber\\
\end{eqnarray}
and the polarization of the pseudo-spin formed by Majorana zero modes is
\begin{eqnarray}
s &=& |\psi_1(t)|^2-|\psi_0(t)|^2=\cos(\omega_{\rm M} t),
\end{eqnarray}
with the initial state $\psi_0 = 0$ and $\psi_1 = 1$ and oscillation angular frequency
\begin{eqnarray}\label{eq:LZSfrequency}
\omega_{\rm M}=2\omega_{\rm LZS} =2\delta J_{0}(2z)=2\delta J_{0}\left(\frac{2E_{\rm M}}{eV_0}\right).
\end{eqnarray}

\section{Josephson Radiation and Bessel oscillations}\label{appendix:B}
As shown in section \ref{II} of the main text, the equation of motion for the Josephson phase is given as
\begin{eqnarray}\label{eq:RCSJ2}
\frac{\hbar C \ddot \theta} {2eI} + \frac{{\rm }\hbar \dot \theta}{2eRI} +I_{\rm 1}\sin{\theta} + I_{\rm 2}\langle i\gamma_{\rm L} \gamma_{\rm R}\rangle \sin{\frac{\theta}{2}}= 1 ,
\end{eqnarray}
which is a self-consistent equation for $\theta$ since the time evolution of the Majorana two-level system has been determined by Eq. (\ref{eq:LZSfrequency}) as,
\begin{eqnarray}\label{eq:sew}
\langle i\gamma_{\rm L} \gamma_{\rm R} \rangle \approx s_1 e^{i\omega_{\rm M}t}+s_1^* e^{-i\omega_{\rm M}t},
\end{eqnarray}
where $s_1$ is a constant decided by the initial condition.
The Josephson radiation $V=2e\dot \theta$ can be obtained by solving Eq.~(\ref{eq:RCSJ2}). Since it is a second-differential equation, we try to solve it with the Fourier transformation.
\begin{eqnarray}
v(\omega)&=&\int_{-\infty}^{+\infty}V(t)e^{-i\omega t}\mathrm{d}t.
\end{eqnarray}
where $v(\omega)$ is the amplitude in frequency domain $\omega$. The pure real function $V(t)$ requires that $v(-\omega)=v^*(\omega)$.
Assuming that $V(t)$ has a discrete spectrum, we can write down the expansion in the form of
\begin{equation}
v(\omega) =V_0\delta(\omega)+\frac{1}{2}\sum_{n}\big[V_n\delta(\omega-\omega_n)+V_n^*\delta(\omega+\omega_n)\big],
\end{equation}
where $\omega_n>0$ represents the peaks in the radiation spectrum. The factor $1/2$ makes $|V_n|$ directly the measured strength of angular frequency $\omega_n$ and arg$(V_n)$ is the phase which also is convenient for later calculation. With this simplified expansion, we can explicitly express the inverse Fourier transformation in the form of
\begin{eqnarray} \label{eq:Vew}
V(t)&=&V_0+\frac{1}{2}\sum_{n}\left[V_ne^{i\omega_nt}+V_n^*e^{-i\omega_nt}\right] \nonumber\\
&=&V_0+\sum_{n}\big[a_n\cos(\omega_nt)+b_n\sin(\omega_nt)\big],
\end{eqnarray}
where we define $a_n\equiv (V_n+V_n^*)/2$, $b_n\equiv i(V_n-V_n^*)/2$. Now we write down the expression for the Josephson phase $\theta$ with Fourier transformation. The integration of the second Josephson relation ${\rm d}\theta/{\rm d} t=2eV$ using Eq.~(\ref{eq:Vew}) gives
\begin{equation} \label{eq:theta_expand}
\theta=\theta_0+2eV_0t+2e\sum_n\left( \frac{a_n}{\omega_n}\sin\omega_nt-\frac{b_n}{\omega_n}\cos\omega_nt \right),
\end{equation}
where the constant $\theta_0$ is decided by the initial condition. This is the Eq.~(\ref{theta}) of the main text. For convenience, we will use $\omega_{\rm J}=2eV_0$ to simplify the factor.
We put the expression for $\theta$ back into the sine functions of Eq.~(\ref{eq:RCSJ2}), and try to obtain an explicit expansion.
For this purpose, a mathematically convenient method is to write the sine functions in the exponential form $\sin\theta=i(e^{-i\theta}-e^{i\theta})/2$. Then we try to expand the exponential functions with Bessel functions. For the first term, we substitute Eq.~(\ref{eq:theta_expand}) in it and obtain
\begin{equation}
e^{i\theta}=e^{i(\omega_{\rm J}t+\theta_0)}\prod_n e^{i\frac{2e a_n}{\omega_n}\sin\omega_nt}e^{-i\frac{2e b_n}{\omega_n}\cos\omega_nt}.
\end{equation}
We define $A_n\equiv2e a_n/(\omega_n)$ and $B_n\equiv2e b_n/(\omega_n)$. They also can be expressed by $V_n$ as
\begin{equation} \label{eq:AnBn}
A_{n} = \frac{e}{\omega_n}(V_n+V_n^*),\quad B_n=\frac{ie}{\omega_n}(V_n-V_n^*).
\end{equation}
Using Jacobi-Anger expansion $e^{iz \sin x} = \sum_{n = -\infty}^{+\infty} J_n (z) e^{i n x}$ and $e^{iz \cos x} = \sum_{n = -\infty}^{+\infty} i^n J_n (z) e^{i n x}$, we arrive at
\begin{equation} \label{eq:eitheta}
e^{i\theta}=e^{i(\omega_{\rm J}t+\theta_0)}\prod_n \sum_{p,q=-\infty}^{\infty}i^qJ_p(A_n)J_q(-B_n)e^{i(p+q)\omega_nt}.
\end{equation}

Up to now, all the formulas are exact.~To obtain analytical solutions, now we begin to make approximations. Using the fact that $J_0\gg J_1\gg\cdots$, we can discard most of the terms in this polymerization of Eq.~(\ref{eq:eitheta}). We keep the zero-order term just containing $J_0$ and the first-order term just containing one $J_{\pm1}$ with all other factors purely $J_0$.
Then the formula is simplified to
\begin{equation}
e^{i\theta}\approx e^{i(\omega_{\rm J}t+\theta_0)} \left[j_0+\sum_n(j_ne^{i\omega_nt}-j_n^*e^{-i\omega_nt})\right],
\end{equation}
where for simplicity we have defined
\begin{eqnarray}\label{eq:j}
j_0 &=& \prod_n J_0(A_n)J_0(B_n) ,\nonumber\\
j_n &=& j_0\frac{J_1(A_n)J_0(B_n)-iJ_0(A_n)J_1(B_n)}{J_0(A_n)J_0(B_n)}.
\end{eqnarray}
Here we have used the relation $J_{-1}(x)=J_1(-x)=-J_1(x)$ and $J_0(-x)=J_0(x)$.
We plug it back to the sine function and get
\begin{equation} \label{eq:sinew}
\sin\theta \approx -\frac{i}{2}e^{i(\omega_{\rm J}t+\theta_0)} \left[j_0+\sum_n(j_ne^{i\omega_nt}-j_n^*e^{-i\omega_nt})\right]+\mathrm{h.c.}.
\end{equation}
In the same way, we can expand the $\sin \theta/2$ term, and obtain the expression up to the order of $J_1$ as,
\begin{equation}\label{eq:sin2ew}
\sin{\frac{\theta}{2}}\approx-\frac{i}{2}e^{i\left(\frac{\omega_{\rm J}}{2}t+\frac{\theta_0}{2}\right)}\left[l_0+\sum_n(l_ne^{i\omega_nt}-l_n^*e^{-i\omega_nt})\right]+\mathrm{h.c.},
\end{equation}
where for simplicity we have defined
\begin{eqnarray}\label{eq:l}
l_0 &=& \prod_n J_0(\frac{A_n}{2})J_0(\frac{B_n}{2}), \nonumber\\
l_n &=& l_0\frac{J_1(\frac{A_n}{2})J_0(\frac{B_n}{2})-iJ_0(\frac{A_n}{2})J_1(\frac{B_n}{2})}{J_0(\frac{A_n}{2})J_0(\frac{B_n}{2})}.
\end{eqnarray}
Now we look at other terms in Eq.~(\ref{eq:RCSJ2}). The time derivative of the voltage is written as
\begin{eqnarray} \label{eq:dVew}
\frac{{\rm d} V}{{\rm d} t} &=&\frac{1}{2} \sum_n\left(i\omega_nV_n e^{i\omega_nt}-i\omega_nV_n^* e^{-i\omega_nt}\right).
\end{eqnarray}
Plugging Eqs.~(\ref{eq:sew}), (\ref{eq:Vew}) and (\ref{eq:sinew})-(\ref{eq:dVew}) into the Eq.~(\ref{eq:RCSJ2}), we find the expansion series of,
%\begin{eqnarray}\label{eq:sum_expo}
\begin{equation}\label{eq:sum_expo}
\begin{aligned}
&2\left(\frac{V_0}{R}-I\right) \\
&=iI_\mathrm{c1}j_0e^{i(\omega_{\rm J}t+\theta_0)} \\
&+iI_\mathrm{c2}l_0s_1 e^{i\left[(\frac{\omega_{\rm J}}{2}+\omega_{\rm M})t+\frac{\theta_0}{2}\right]}+iI_\mathrm{c2}l_0s_1^* e^{i\left[(\frac{\omega_{\rm J}}{2}-\omega_{\rm M})t+\frac{\theta_0}{2}\right]} \\
&+\sum_n \Bigg\{{- \left(\frac{1}{R}+i\omega_nC\right)V_n e^{i\omega_nt}} \\
&+iI_\mathrm{c1}j_ne^{i[(\omega_{\rm J}+\omega_n)t+\theta_0]}-iI_\mathrm{c1}j_n^*e^{i[(\omega_{\rm J}-\omega_n)t+\theta_0]} \\
&+iI_\mathrm{c2}s_1l_ne^{i\left[(\frac{\omega_{\rm J}}{2}+\omega_{\rm M}+\omega_n)t+\frac{\theta_0}{2}\right]}-iI_\mathrm{c2}s_1l_n^*e^{i\left[(\frac{\omega_{\rm J}}{2}+\omega_{\rm M}-\omega_n)t+\frac{\theta_0}{2}\right]}\\
&+iI_\mathrm{c2}s_1^*l_ne^{i\left[(\frac{\omega_{\rm J}}{2}-\omega_{\rm M}+\omega_n)t+\frac{\theta_0}{2}\right]}\\
&-iI_\mathrm{c2}s_1^*l_n^*e^{i\left[(\frac{\omega_{\rm J}}{2}-\omega_{\rm M}-\omega_n)t+\frac{\theta_0}{2}\right]} \Bigg\} \quad +\mathrm{h.c.},
\end{aligned}
\end{equation}
%\end{eqnarray}
where we have multiplied the equation by 2 and rearrange $V_0$ %and $V_n$
to the left hand side of the equation. Using the fact that two Fourier eigenfunctions are orthogonal,
\begin{eqnarray}
\frac{1}{2\pi}\int_{-\infty}^{+\infty}e^{-i\omega_mt}V_ne^{i\omega_nt}\mathrm{d}t &=&V_m \delta(\omega_m-\omega_n),
\end{eqnarray}
we can calculate $V_n$ by multiplying both sides of Eq.~(\ref{eq:sum_expo}) with $e^{-i \omega_m t}$
and then integrate over the full-time range. We find that $V_n \neq 0$ only for the discrete frequencies of
\begin{eqnarray}
\omega_n = p\frac{\omega_{\rm J}}{2}+q \omega_{\rm M} \quad(p,q,\frac{p+q}{2}\in Z, \omega_n>0),
\end{eqnarray}
where the lowest order radiation frequencies are explicitly written as
\begin{eqnarray}
\omega_1=\frac{\omega_{\rm J}}{2}-\omega_{\rm M}, \quad\omega_2=\frac{\omega_{\rm J}}{2}+\omega_{\rm M}, \quad \omega_3=\omega_{\rm J}.
\end{eqnarray}
Recalling Eq.~(\ref{eq:LZSfrequency}), finally we get the radiation spectra
\begin{eqnarray}\label{eq:MRadiap}
&&f_1 = \frac{f_{\rm J}}{2}- {\frac{\delta}{\pi} J_{0}\left(\frac{2E_{\rm M}}{eV_0}\right)},\quad \nonumber \\
&&f_2 = \frac{f_{\rm J}}{2}+ {\frac{\delta}{\pi} J_{0}\left(\frac{2E_{\rm M}}{eV_0}\right)},\nonumber \\
&&f_3= f_{\rm J},
\end{eqnarray}\label{eq:MRadia}
which is the Eq.~(\ref{eq:MRadi}) of the main text.

Now we consider the results shown in Fig.~\ref{fig:four} in which the junction parameters are taken in the regime of large $\delta$. In this regime, the Bessel function $J_0(z)$ is approximately equal to $\sqrt{2/\pi z}\rm{cos}(z-\pi/4)$ since $z$ is large. Then for small dc voltage, $f^n_{\rm M\pm}$ described by Eq.~(\ref{eq:higherbessel}) can be approximately written as
\begin{equation}\label{eq:simmlarity}
f^n_{\rm M \pm}=\frac{(2n-1)eV_0}{2\pi}\pm\frac{\delta}{\pi}\sqrt{\frac{eV_0}{\pi E_{\rm M}}}{\rm cos}(\frac{2E_{\rm M}}{eV_0}-\frac{\pi}{4}).
\end{equation}
With this expression, we can obtain $f$ and $V_0$ corresponding to the $m$th intersection point of $f_{\rm M+}^n$ and $f_{\rm M-}^n$ as
\begin{eqnarray}\label{eq:f and V0}
f/(E_M/2\pi)_{m,n}=\frac{4n-2}{m\pi-\pi/4}, \\ V_0/(E_M/2e)_{m,n}=\frac{4}{m\pi-\pi/4}\nonumber.
\end{eqnarray}
We define a new parameter $l=m-n$. Then the black lines in Fig.~\ref{fig:four}(b) connect points $(f,V_0)$ described by Eq.~(\ref{eq:f and V0}) with same $l$. Replace $n$ in Eq.~(\ref{eq:f and V0}) with $l$ and we can find
\begin{eqnarray}\label{eq:f and V02}
[\frac f {(E_M/2\pi)}]_{m,l}&&=\frac{4m-4l-2}{m\pi-\pi/4}\\
&&=\frac 4 \pi-(l+\frac 1 4)\frac{4}{m\pi-\pi/4}\nonumber\\
&&=\frac 4 \pi-(l+\frac 1 4)[\frac{V_0}{(E_{\rm M}/2e)}]_{m,l},\nonumber
\end{eqnarray}
which is the Eq.~(\ref{eq:mine}) of the main text. The expression gives the black lines in Fig.~\ref{fig:four}(b).

Finally, we examine the radiation intensity for these lowest order radiation frequencies and take the approximation that the contributions from $\omega_1$, $\omega_2$, and $\omega_3$ dominant the summation in Eq.~(\ref{eq:sum_expo}). Then for these three frequencies, we obtain three equations respectively,
\begin{equation} \label{eq:V123}
\begin{aligned}
\left(\frac{1}{R}+i\omega_1C\right)V_1&=iI_\mathrm{c2}l_0s_1^*e^{i\frac{\theta_0}{2}}-iI_\mathrm{c1}j_2^*e^{i\theta_0}, \\
\left(\frac{1}{R}+i\omega_2C\right)V_2&=iI_\mathrm{c2}l_0s_1e^{i\frac{\theta_0}{2}}-iI_\mathrm{c1}j_1^*e^{i\theta_0}, \\
\left(\frac{1}{R}+i\omega_3C\right)V_3&=iI_\mathrm{c1}j_0e^{i\theta_0}+iI_\mathrm{c2}s_1l_1e^{i\frac{\theta_0}{2}}+iI_\mathrm{c2}s_1^*l_2e^{i\frac{\theta_0}{2}} ,\\
\end{aligned}
\end{equation}
where the functions $j_{0,1}$ and $l_{0,1,2}$ are functions of $A_{1,2,3}$ and $B_{1,2,3}$ as defined in Eqs.~(\ref{eq:j}) ad (\ref{eq:l}), with the parameters $A_n$, $B_n$ in Eq.~(\ref{eq:AnBn}).
In the regime of $I\gg I_{\rm c1}+I_{\rm c2}$, we have $A_n,B_n \ll 1$. As the lowest order approximation we take $J_0 \approx 1$ and neglect all higher-order Bessel functions in Eqs.~(\ref{eq:j}) and (\ref{eq:l}). We obtain
\begin{eqnarray}
j_0=l_0=1, \quad
j_n=l_n=0 .
\end{eqnarray}
We can go further to the first-order approximation by including the Bessel function $J_1$ and taken $J_1(x)=x/2$ in Eqs.~(\ref{eq:j}) and (\ref{eq:l}). Then we obtain
\begin{equation}
j_0=l_0=1,\quad j_n=\frac {A_n-iB_n} 2=\frac{e}{\omega_n}V_n,\quad l_n=\frac{e}{2\omega_n}V_n.
\end{equation}
Plugging this expression back into the Eq.~(\ref{eq:V123}), we obtain the solution for $V_n$,
\begin{eqnarray}
V_1 &=& \frac{ iI_\mathrm{c2}{s_1^*}e^{i\frac{\theta_0}{2}}\left(\frac{1}{R}-i\omega_2C\right)- \frac{eI_{\rm c1}I_{\rm c2}}{\omega_2} s_1^* e^{i\frac{\theta_0}{2}} }
{ \left(\frac{1}{R}+i\omega_1C\right)\left(\frac{1}{R}-i\omega_2C\right)-\frac{e^2 I_{\rm c1}^2}{\omega_1\omega_2} }, \\
V_2&=& \frac{ iI_\mathrm{c2}s_1 e^{i\frac{\theta_0}{2}}\left(\frac{1}{R}-i\omega_1C\right)- \frac{eI_{\rm c1}I_{\rm c2}}{\omega_1}s_1e^{i\frac{\theta_0}{2}} }
{ \left(\frac{1}{R}-i\omega_1C\right)\left(\frac{1}{R}+i\omega_2C\right)-\frac{e^2 I_{\rm c1}^2}{\omega_1\omega_2} },\nonumber \\
V_3&=&\frac{iI_\mathrm{c1}e^{i\theta_0}+iI_\mathrm{c2}s_1\frac{e}{2\omega_1}V_1 e^{i\frac{\theta_0}{2}}+iI_\mathrm{c2}s_1^*\frac{e}{2\omega_2} V_2 e^{i\frac{\theta_0}{2}}}
{\frac{1}{R}+i\omega_{3} C}.\nonumber
\end{eqnarray}
This gives the final analytical formula for the radiation intensity of the topological Josephson junction which includes the quantum dynamics of the two-level system formed by Majorana zero modes.

\bibliography{reference}

\end{document}